\documentclass[12pt,preprint]{emulateapj}

\usepackage{amssymb}
\usepackage{natbib}
\usepackage{graphicx}

\def\LCDM{$\Lambda\mbox{CDM}$  }

\def\beqn{\vspace{2mm} \begin{eqnarray}}
\def\eeqn{\vspace{2mm} \end{eqnarray}}
\newcounter{parentequation}

\def\cm{{\rm\thinspace cm}}
     
\def\g{{\rm\thinspace g}}

\def\Mpc{{\rm\thinspace Mpc\ }}

\slugcomment{Not to appear in Nonlearned J., 45.}

\shorttitle{Cosmological Ellipticals}
\shortauthors{Naab et al.}

\begin{document}


\title{Formation of early-type galaxies from cosmological initial conditions}

\author{Thorsten Naab$^{1,2}$, Peter H. Johansson$^2$, Jeremiah P. Ostriker$^{2,3}$ \& 
George Efstathiou$^{2}$} 
\affil{$^1$ Universit\"ats-Sternwarte M\"unchen, Scheinerstr.\ 1, D-81679 M\"unchen, Germany; \texttt{naab@usm.lmu.de}\\
$^2$ Institute of Astronomy, Madingley Road, Cambridge CB3 0HA, UK \\
$^3$ Department of Astrophysics, Peyton Hall, Princeton, USA}



\begin{abstract}
We describe high resolution Smoothed Particle Hydrodynamics (SPH)
simulations of three approximately $M_*$ field galaxies starting 
from \LCDM initial
conditions. The simulations are made intentionally simple, and include
photoionization, cooling of the intergalactic medium, and star formation 
but not feedback from AGN or supernovae. All of the galaxies undergo an 
initial burst of star
formation at $z \approx 5$, accompanied by the formation of a bubble of
heated gas. Two out of three galaxies show early-type properties at present 
whereas only one of them experienced a major merger.  
Heating from shocks and -PdV work dominates over cooling so that 
for most of the gas the temperature is an increasing function of time.  
By $z \approx 1$ a significant fraction of the final stellar mass is in 
place and the spectral energy distribution resembles those of observed 
massive red galaxies. The galaxies have grown from $z=1 \rightarrow 0$ on 
average by 25\% in mass and in size by gas poor (dry) stellar mergers. 
By the present day, the simulated galaxies are old ($\approx
10 \;{\rm Gyrs}$), kinematically hot stellar systems surrounded by hot 
gaseous haloes. Stars dominate the mass of the galaxies up to 
$\approx 4$ effective radii ($\approx 10$ kpc). Kinematic and most 
photometric properties are in good agreement with those of 
observed elliptical galaxies. The galaxy with a major merger develops 
a counter-rotating core.  Our simulations 
show that realistic intermediate mass giant 
elliptical galaxies with plausible formation histories can be formed 
from \LCDM initial conditions even without requiring recent major mergers or 
feedback from supernovae or AGN.
\end{abstract}


\keywords{galaxy formation: general --- galaxy formation: elliptical --- methods: numerical}



\section{Introduction}

There are many puzzles encountered in understanding the formation and
evolution of elliptical galaxies and the spheroidal components of
spiral galaxies. On the one hand a naive reading of the
hierarchical theory of structure formation in a $\Lambda$CDM universe
would argue that, since massive halos form later than less massive
ones, massive ellipticals which reside in the centers of these
massive halos should also form late.  But there is strong observational 
evidence that old, massive, red and metal rich proto-ellipticals are 
already in place at $z=2-3$ and that
present day early-type galaxies formed most of their stars well before a 
redshift $z=1$ \citep{1973ApJ...179..427S,2000ApJ...536L..77B,2005ApJ...633..174T,
2005ApJ...631..145V}. We also know from evidence
dating to the 1970's that current rates of star formation in these
systems are quite low, the rates increasing sharply into the past as
$z^1$ \citep{1978ApJ...219...18B,1980ApJ...236..351D,1995ApJ...439...47R}.  
Further, with regard 
to the evolution of this population,
there appears to be a significant increase in the total stellar mass in these old
objects from $z=1$ to the present \citep{2004ApJ...608..752B,2004ApJ...608..742D,
2005ApJ...620..564C,2005astro.ph..6044F}, 
but this cannot easily be accounted for by the fading of
younger, bluer star forming galaxies present at $ z =1$ \citep{2004ApJ...608..742D,
2005astro.ph..6044F}.  This has
led to the plausible and popular idea that `dry merging' (i.e. merging of 
predominantly stellar systems) among elliptical
systems \citep{2005ApJ...627L..25T,2005AJ....130.2647V,2006astro.ph..2038B,
2006ApJ...640..241B} pushes more and more of the mass over the observational
cutoff to provide an increased number of 0.5-2$L_*$ galaxies. 
  
This attractive idea gives reasonable explanations for some observed correlations 
of the mass of ellipticals and their kinematics, sizes or isophotal shape 
\citep{2005MNRAS.359.1379K,2006ApJ...636L..81N,2006MNRAS.369.1081B}. However, noting that 
there are tight relations among luminosity, color, age and metalicity, 
puts severe limits on the amount of late dry merging that could acceptably 
occur without destroying the aforementioned constraints.

On the theoretical side it is now possible to perform accurate high 
resolution simulations of the gravitational evolution of the dark matter
distribution \citep{1998ApJ...499L...5M,2005Natur.435..629S}.
In contrast, the numerical simulation of galaxy formation, including a
hydrodynamic treatment of the baryonic component, is still in its
infancy. Very few high resolution simulations from realistic
cosmological initial conditions have been done so far and 
most of these have concentrated on the formation
of disc galaxies rather than early-type spheroidal systems 
\citep{2003ApJ...596...47S,2003ApJ...597...21A,2004ApJ...607..688G,2004ApJ...606...32R}.

This is surprising since spheroidal systems are of interest in their
own right, as they contain more than half of the total stellar mass in
the local universe \citep{1998ApJ...503..518F}. The most massive
galaxies known,  the giant ellipticals, 
are spheroidal systems which  predominantly consist of
old stars (see e.g. \citealp{2005ApJ...621..673T}) and so must have
formed at high redshift. They are therefore likely to be good probes
of galaxy assembly, star formation and metal enrichment in the early
Universe. 
\begin{figure}
  \centering
  \includegraphics[width=8cm]{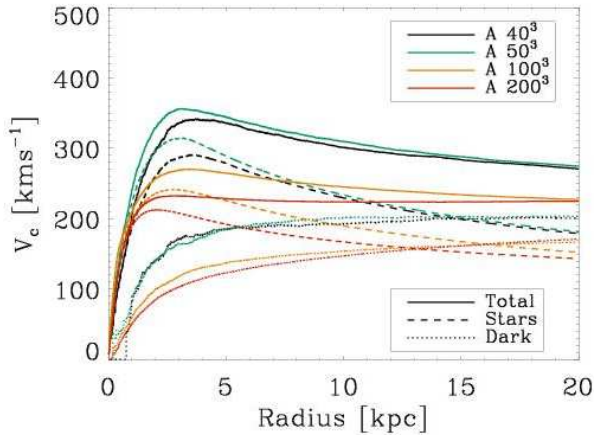}
  \caption{Circular velocity curves for galaxy A at four different numerical resolutions: 
$40^3$, $50^3$, $100^3$, and $200^3$ SPH particles and collisionless
    dark matter particles, respectively. Note how the rotation curves become increasingly flat as 
the resolution increases. \label{rotcurve_haloA}}
\end{figure}
\begin{figure}
  \centering
  \includegraphics[width=8cm]{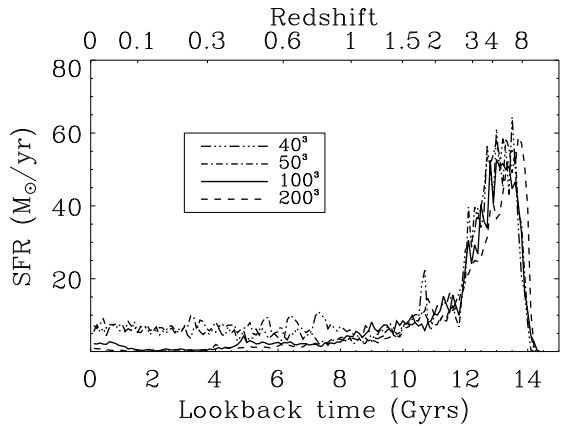}
  \caption{Star formation rate (SFR) histories, computed from stellar ages, 
  of galaxy A versus lookback time at four different numerical resolutions: 
$40^3$, $50^3$, $100^3$, and $200^3$ SPH particles and collisionless
    dark matter particles, respectively. There is a strong trend that
    the low redshift star formation rate is reduced in higher
    resolution simulations. 
 \label{sfr}}
\end{figure}

Most numerical work on early-type galaxy formation has used either
very idealized initial conditions 
\citep{2004MNRAS.347..740K}, or had insufficient
spatial and mass resolution to resolve the internal structures of
galaxies \citep{2003MNRAS.346..135K,2004ApJ...601L.131S}. An exception
is the simulation discussed by \citet{2003ApJ...590..619M}. These
authors used an SPH simulation that included feedback from supernovae to
follow the formation of a single spheroidal galaxy self-consistently
from CDM initial conditions. The spatial resolution of this simulation
was high enough to resolve the region within an effective radius for 
a typical real  elliptical galaxy. However, the final stellar system 
formed in this simulation was far too dense, with an effective radius about an 
order of magnitude smaller than real elliptical galaxies of the same brightness.
\citet{2003ApJ...590..619M} speculate that this discrepancy may be a 
consequence of their star formation and feedback algorithm and that
it might be possible to produce less concentrated systems if more aggressive
stellar feedback were implemented to prevent star formation in high density
sub-units at high redshift.

In this paper we present high resolution hydrodynamical simulations 
based on cosmological initial conditions
admitting, intentionally, only bare-bones prescriptions for the physics
involved (e.g. no ``feedback'' from supernovae or AGN), to see if some resolution of these
paradoxes can be derived.  The rationale for
simplyfing the simulations is straightfoward. As we will show in Section \ref{MASSIVE}, 
at least $100^3$ gas particles are required for
hydrodynamical simulations to `converge', making them very expensive to
run. With present generation computers it is impossible to run a large
ensemble of simulations of this size to properly explore a huge
parameter space. Our point of view, therefore, is to keep the
physics of the simulations as simple as possible and to get an
understanding of the behaviour of a simplified problem before
investigating additional complexities such as supernova and AGN
feedback. Our goalin this paper, therefore, is to investigate
the formation of a number of massive ($\approx M_*$) isolated galaxies
starting from realistic initial conditions and to see how variable
the final systems and  whether they resemble real galaxies. 

What we find can be summarized simply.  The
initial cooling and collection of cold gas from infalling smaller
scale perturbations happens very rapidly and easily within the most massive
systems leading to a very rapid burst of star formation
beginning at $z \sim 6$ and then falling off exponentially on a
time-scale of roughly 1.5Gyrs. 
This phase is terminated as the star forming
region is enveloped in an expanding hot bubble which prevents new,
infalling cold gas mass elements from reaching the central regions.
This early phase - reminiscent in some respects to a modernized version
of the ``monolithic collapse'' picture of galaxy formation, prominent
in the 1960's through 1980's \citep{1974MNRAS.169..229L,1982MNRAS.201..939V}- 
produces a sequence of objects with
effective radii of 1-2kpc, which might satisfy the tight relations observed
among the red metal rich old cores of elliptical galaxies.  

For simulations with early-type properties at $z=0$ stellar accretion 
or mergers (the choice of the appropriate term for the process is 
arbitrary and dependent on the relative mass of
the infalling stellar objects) add to the growing stellar envelope of
relatively blue, old and metal poor stars.
This accounts for the growth in size seen for ellipticals as well as the 
growth in total mass in the time frame $z=1 \rightarrow 0$.  Furthermore, since this
assembly of added stellar mass is not accompanied by much in situ
star formation (the coterminously added hot and cold gas being simply added to the
expanding hot bubble surrounding the central galaxy) the growth of the
stellar population occurs without the presence of young stars.  Existing work has 
shown that the tightness of the elliptical color-magnitude relation puts 
strong constraints on dry merger scenarios 
\citep{1998MNRAS.299.1193B,2005MNRAS.360...60K}. But we find that in our 
simulations that minor mergers or accretion events do not typically add much stellar mass to
the central region, because of the relatively large angular momentum of the
infalling systems - and thus the central, tight color-magnitude relations
are not expected to be overly perturbed even if as much as 40\% of the
total final mass is added during this phase.  As an additional byproduct of
this assembly scenario, we find it readily understandable how so much of the
stellar mass in these systems resides in regions of such low density,
where star formation would always have been difficult to contemplate.
The stars seen at $R \sim R_e$ were not born there, but rather were formed in
the central regions of much lower mass systems that have been accreted
and shredded in the envelopes of the giant elliptical galaxies.

\begin{table*}
\caption{Properties of galaxy A at $r < r_{\mathbf{vir}}$ }             
\label{A_global}      
\centering          
\begin{tabular}{c| c c c c c c | c c c c c c}     
\hline\hline       
Resolution & $M_{\mathbf{vir}}$ \tablenotemark{(a)} & $M_{\mathbf{stars}}$ &$M_{\mathbf{gas}}$ & $M_{\mathbf{dark}}$ &  $r_{\mathbf{vir}}$ \tablenotemark{(b)}&  $v_{\mathbf{max}}$  \tablenotemark{(c)}&  $m_{\mathbf{stars}}$ \tablenotemark{(d)} & $m_{\mathbf{dark}}$& $\epsilon_{\mathbf{stars}}$ \tablenotemark{(e)}& $\epsilon_{\mathbf{dark}}$  \\ 
\hline                    
  $40^3$  & 242 &  25.3 & 20.2 & 197 & 434 & 341 &  161 & 1288 & 0.625& 1.3\\  
  $50^3$  & 241 &  27.0 & 20.1 & 193 & 433 & 356 &  82.5 & 659  & 0.5  & 1.0\\
  $100^3$ & 230 &  23.8 & 20.9 & 184 & 427 & 270 &  10.5 &  82   &0.25 & 0.5\\
  $200^3$ & 225 &  24.9 & 18.4 & 182 & 424 & 232 &  1.3 & 10.3  &0.125& 0.25 \\
\hline                  
\end{tabular}
\tablecomments{(a) Total masses $M$ in$10^{10}M_{\odot}$; (b) Virial radius in kpc; 
(c) Maximum circular velocity in km/s;
(d) Particle masses $m$ in $10^{5}M_{\odot}$; (e) Gravitational softening lengths in kpc}

\end{table*}

\begin{table}
\caption{Properties of galaxy A at $r < 30$kpc}             
\label{A_30}      
\centering          
\begin{tabular}{c| c c c}     
\hline\hline       
Resolution & $M_{\mathbf{stars}}$ \tablenotemark{(a)}& $M_{\mathbf{gas}}$ &$M_{\mathbf{dark}}$  \\ 
\hline                    
  $40^3$  & 16.3 &  1.0 & 27.7 \\  
  $50^3$  & 16.6 &  1.0 & 28.7 \\
  $100^3$ & 12.2 &  0.6 & 20.4 \\
  $200^3$ & 11.8 &  0.5 & 23.1 \\
\hline                  
\end{tabular}
\tablecomments{(a) Total masses $M$ in$10^{10}M_{\odot}$}
\end{table}

\section{Initial conditions and simulations}
The initial conditions of the \LCDM simulation assumed scale-invariant 
adiabatic fluctuations. The post-recombination 
power spectrum for the CDM cosmology was based on
the parameterization of \citet{1992MNRAS.258P...1E} with $\Gamma$=0.2.
We use a WMAP1 \citep{2003ApJS..148..175S} cosmology with a slightly lower 
Hubble parameter of $h=0.65$ ($\equiv H_{0}$=100$h$ kms$^{-1}$Mpc$^{-1}$) 
with $\sigma_8$=0.86, $f_{b}= \Omega_b/\Omega_m$=0.2, $\Omega_0$=0.3, and 
$\Lambda_0$=0.7. 

In a low resolution, dark matter simulation (128$^3$ dark matter 
particles with a box size of  $L_{box}$ = 50 \Mpc using the (AP$^3$M) N-body code of
\cite{1991ApJ...368L..23C}). We selected virialized 
halos with masses in the range $7\times 10^{11}M_{\odot} < M_{halo} 
< 3 \times 10^{12}M_{\odot}$ in a low density environment such 
that the nearest halo with $M_{halo} > 2 \times$ 10$^{11}M_{\odot}$ is more than
one \Mpc/h away. To re-simulate the target halos at high resolution we 
increased the particle number to $100^{3}$ gas and dark matter particles within
a cubic volume at redshift $z=24$ containing all particles that end up
within the virialized region (conservatively we assumed a fixed radius of $0.5 \ \rm Mpc$) 
of the halos at $z=0$. From this sample we have selected three systems 
with final stellar masses of $\approx M_*$ for elliptical galaxies in the range of 
$9.6 \times 10^{10}M_{\odot} < M < 1.6 \times 10^{11}M_{\odot}$ 
\citep{2001MNRAS.326..255C,2001ApJ...560..566K,2003ApJS..149..289B} 
for the cosmology chosen here. For one halo (halo A) the resolution was increased 
to $200^3$ particles. Additional short wavelength perturbations were 
included to account for the missing small-scale power below the Nyquist 
frequency of the low resolution simulation. The tidal forces from particles 
outside the high resolution cube were approximated by increasingly massive 
dark matter particles in 5 nested layers of lower and lower resolution. 
None of the galaxies was contaminated by boundary particles within their virial radii.


The simulations were run with GADGET-2 \citep{2005MNRAS.364.1105S} on COSMOS, 
a shared-memory Altix 3700 with 1.3-GHz Itanium2 processors hosted at the Department of 
Applied Mathematics and Theoretical Physics (Cambridge) and on two similar 
machines at Princeton University and the University Observatory in Munich. 
We used a fixed comoving softening until $z=9$, and after this the 
softening remained fixed at the same values in physical coordinates. 
The simulation parameters for each halo are given in Tab.\ref{all_vir}. 
The amount of stars formed for each gas particle is proportional to the 
mass of the star forming gas component and inversely proportional to the star formation 
timescale $t_{\star}$. This timescale is $t_{\star}=t_{0}^{\star} \sqrt{\rho_{\rm crit}/\rho}$, 
where $t_{0}^{\star}$ is the characteristic star formation timescale which 
we set to $t_{0}^{\star}=1.5 \; \rm Gyr$ \citep{2003MNRAS.339..289S}. Within a factor 
of a few the exact value of $t_{0}^{\star}$ does not change the results.  

The density threshold for the onset of star formation was set 
to $\rho_{crit}= 7 \times 10^{-26} \rm \g \rm \cm^{-3}$ following 
\citet{1993MNRAS.265..271N}. In addition we require an over-density contrast 
of $\Delta > 55.7$ for the onset of star formation in order 
to avoid spurious star formation at high redshift. 
We included an uniform UV background radiation field of a modified 
\citet{1996ApJ...461...20H} spectrum, where reionization takes place at 
$z \simeq 6$ \citep{1999ApJ...511..521D} and the intensity of the UV background
field peaks at $z\simeq 2-3$. Following \citet{1996ApJS..105...19K} the radiative 
cooling and heating rates were computed assuming that the gas is optically
thin and in ionization equilibrium. The self-shielding of star-forming clouds
from the UV flux (e.g. \citealp{2000ApJ...537..578S,2002ApJ...568L..71Z})
was not included in the simulations. The abundances
of the different ionic hydrogen and helium species were then computed 
by solving the network of equilibrium equations self-consistently for a 
specified value of the UV background radiation field. The late star formation
history is relatively insensitive to minor changes in the UV background field, 
whereas an increase in the UV background at high redshift might
soften the star formation peak. A detailed investigation of the effect
of the UV background on the simulations presented here will be
discussed in a future paper.  

\section{A massive galaxy at z=1}
\label{MASSIVE}
\begin{figure}
\centering 
\includegraphics[width=8cm]{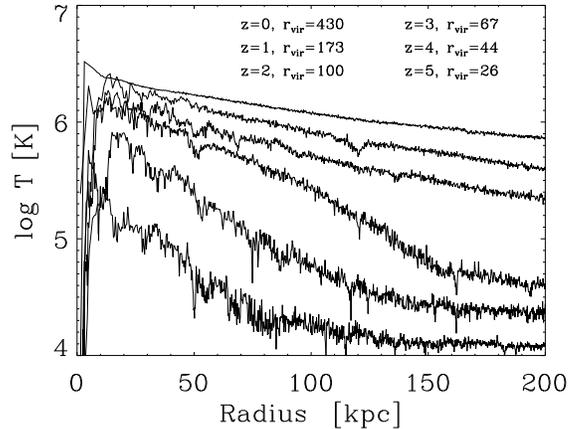}
\caption{Time evolution of the gas temperature profile from $z=5$
  to $z=0$ (from bottom to top) for halo A ($200^3$ resolution). 
The average temperature of the gas is steadily increasing. 
At the end of its initial formation phase at $z\approx2$ the galaxy is surrounded 
by a halo of hot gas heated to the virial temperature.\label{temp_vs_rad}}
\end{figure}
\begin{figure}
\includegraphics[width=8cm]{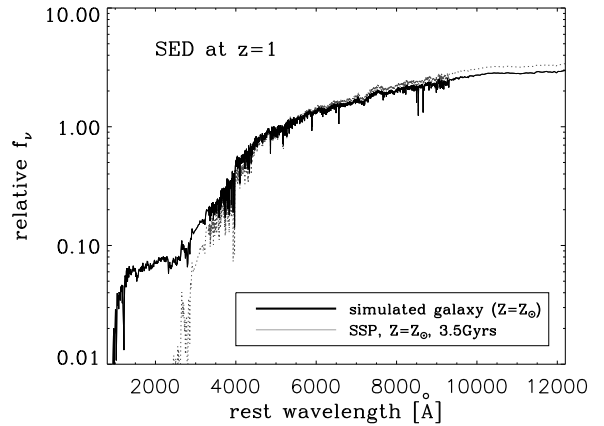}
\caption{Spectral energy distribution (SED) of the galaxy A ($200^3$) at $z=1$ 
(black line) with solar metalicity for all stars. We have assumed that 
stars younger than $10^7$ yrs are obscured. For comparison we show the SED for a 
$3.5\;\rm Gyrs$ old simple stellar population (SSP) 
with solar metalicity (grey dotted line) using the \citet{2003MNRAS.344.1000B}
models with a Salpeter IMF.   
\label{sed}}
\end{figure}

To investigate the effect of numerical resolution on the cosmological simulation 
of an individual galaxy we have simulated our primary example, halo A, at four 
different resolutions 
using $40^3$, $50^3$, $100^3$ and $200^3$ SPH particles and the same number of 
dark matter particles, respectively. The simulation parameters are given in Tab. 
\ref{A_global}. 

The virial radius and the enclosed total mass in baryons and dark matter of the 
galaxy changes weakly with increasing resolution (see Tab. \ref{A_global}). 
However, the amount of baryons concentrated to the central 30 kpc at the 
higher resolutions is reduced by more than 20\% compared to the lowest resolution 
(see Tab. \ref{A_30}). In addition the circular velocity profiles $v^2_c = GM(r)/r$
 become flatter with increasing resolution and the peaks are shifted to smaller radii 
(Fig. \ref{rotcurve_haloA}).

The detailed star formation history of the galaxies also changes 
with numerical resolution. In Fig. \ref{sfr} we compare 
the SFR histories of the $40^3$, $50^3$, $100^3$, and $200^3$
simulations. The numerical resolution has a weak influence on the SFR at
redshifts higher than $z=1$. In all cases the SFRs peak at $z\approx5$ at rates of 
$\approx 60 M_{\odot}/$yr. However, at the $200^3$ resolution the
peak of the SFR is shifted to higher redshifts. At redshifts below
$z=1$ the low resolution galaxies, e.g. $40^3$ and $50^3$
particles, have significantly higher star formation
rates (and gas fractions, see Tab. \ref{A_30}) than  the $100^3$ and 
the $200^3$ simulation which has the lowest present day star formation
rate. The higher star formation rates (and larger final baryon fractions)
of the low resolution simulations are caused by a continuing infall of
cold gas, even at low redshifts. It has been shown by
\citet{2006astro.ph.10051A} and \citet{2006IAUS..235E.214J} that SPH,
especially at low resolution, has a limited abillity to resolve Kelvin-Helmholtz
instabilities: Instead of being destroyed a cold gas cloud moving in a
hot gaseous halo can be artificially stabilized and sink all the way to the
center. In our simulations this artifical effect is suppressed at higher
resolution as the infalling clumpy gas is better resolved and can be
stripped and dispersed in the hot gaseous halo of the host
galaxy.   

A comparison simulation at the $100^3$ resolution with a longer star formation 
time-scale of $t_{0}^{\star}=4.5 \; \rm Gyr$ is similar to the 
$t_{0}^{\star}=1.5 \; \rm Gyr$ simulation at high redshifts but shows significantly 
less star formation at low redshift. 
These results implicate that the resolution has 
a strong effect and the actual choice of the time-scales has a weak effect on the galaxy 
properties. As the present day color of the galaxy is very sensitive to recent star 
formation it is very problematic to use it as a diagnostic tool without considering 
other galaxy properties. 

We have used the highest resolution ($200^3$) simulation to study the
star formation in some more detail. The stars form in clumps of cold 
gas that have collapsed on small scales. As the cold gas is depleted, 
the SFR declines rapidly. The diffuse gas in the forming halo, which does not reside 
in cold sub-clumps, is heated by shocks and forms an expanding bubble of hot gas which 
reaches virial temperature at $2 < z < 3$. As an example we show the
evolution of the gas temperature of halo A in Fig. \ref{temp_vs_rad}.
Although during this phase the cooling time of the hot gas is shorter than 
the Hubble time, the heating time of the infalling gas is even shorter 
\citep{2001ApJ...551..131C,2003MNRAS.345..349B,2005MNRAS.358..168S,2006MNRAS.368....2D}.
After $z=2$ there is an increasing contribution to heating from -PdV
work and entropy addition via weak shocks which becomes dominant towards $z=0$.
In this paper we focus on the assembly history and the observable properties of 
the simulated systems. A full quantitative analysis of the heating processes is 
beyond the scope of this paper and will be presented in Johansson et 
al. (2006, in preparation). 

Galaxy A ($200^3$) has assembled 80\% of its 
final stellar mass by $z=1$ (see Section \ref{ASSEMBLY}). At this time the stellar 
mass within $30$ kpc (we use this 
fiducial value as the stars in ellipticals typically can not be observed at larger radii) 
is $M_{*}= 9.8 \times 10^{10}M_{\odot}$ and the mean age of the stellar 
population is $3.9\;\rm{Gyrs}$.  We constructed the spectral energy 
distribution (SED) of the galaxy at 
$z=1$ using the spectro-photometric stellar evolution models of 
\citep{2003MNRAS.344.1000B} assuming a mean stellar 
metalicity close to solar $Z=Z_{\odot}=0.02$ and a Salpeter IMF 
(Fig. \ref{sed}). Qualitatively the SED does not change for 
small changes in metalicity. We also assumed that recently formed stars 
younger than $10^{7}$ yrs are obscured and do not contribute to the total flux. 
No further corrections for reddening, e.g. for absorption by dust, have been applied. 
The SED looks remarkably similar to rest-frame SEDs of ERO's 
\citep{2004ApJ...600L.131M}. However, in the absence of reddening corrections, 
the redshift corrected observed colors are not quite extreme enough 
at $R-K=4.6$ and $I-K=3.2$. With an observed magnitude of 
$m_{K}=19.3$ and the above red colors the object could alternatively 
be classified as a Very Red Object, see e.g. 
\citep{2004ARA&A..42..477M,2004A&A...421..821V}. 

\begin{figure}
  \centering
  \includegraphics[width=8cm]{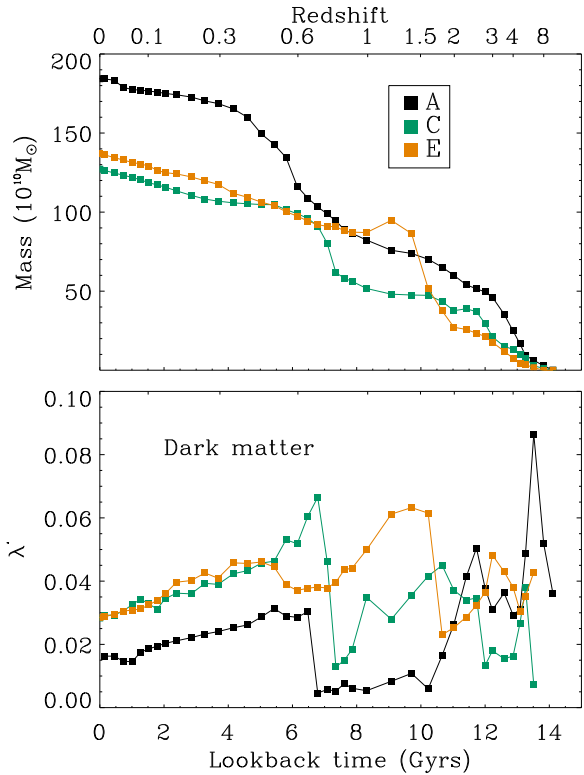}
  \caption{{\it Top}: Dark matter mass accretion history of the three ($100^3$) halos 
within their virial radii. Halo E accretes smoothly after $z=1$. 
{\it Bottom}: Evolution of the halo spin parameter $\lambda^`$. 
The merger events for halos A and C at $z\approx 0.7$ can be clearly 
identified by the sudden increase of the spin. \label{track_all}}
\end{figure}

\begin{figure}
  \centering
  \includegraphics[width=8cm]{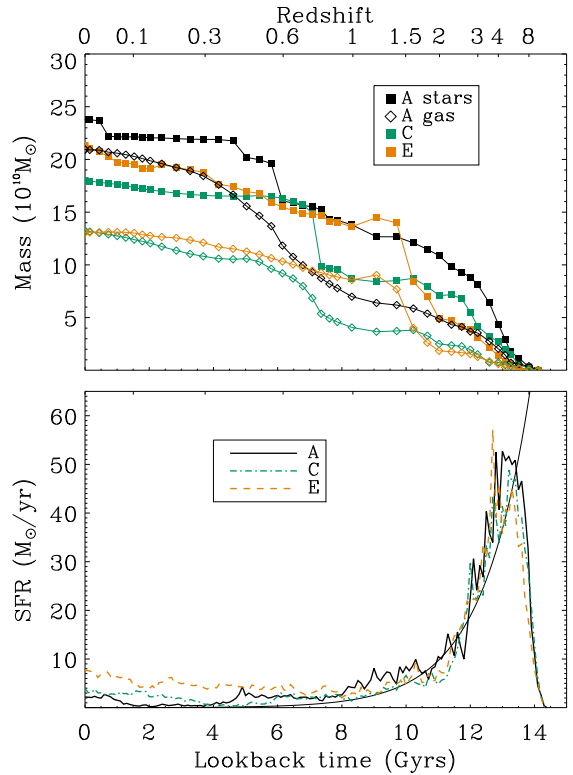}
  \caption{{\it Top}: Baryonic mass accretion history within the virial 
radius of the galaxies subdivided in stars and gas. {\it Bottom}: Star formation 
rate histories computed from stellar ages at $z=0$. The dashed line indicates 
an evolution $ \propto \exp{(-(t-t_0)/\tau)}$  
with $t_0=14.5\;\rm Gyrs$ and $\tau = 1.5\;\rm Gyrs$ for comparison.   
    \label{sfr_all}}
\end{figure}

\begin{figure}
  \centering
  \includegraphics[width=8cm]{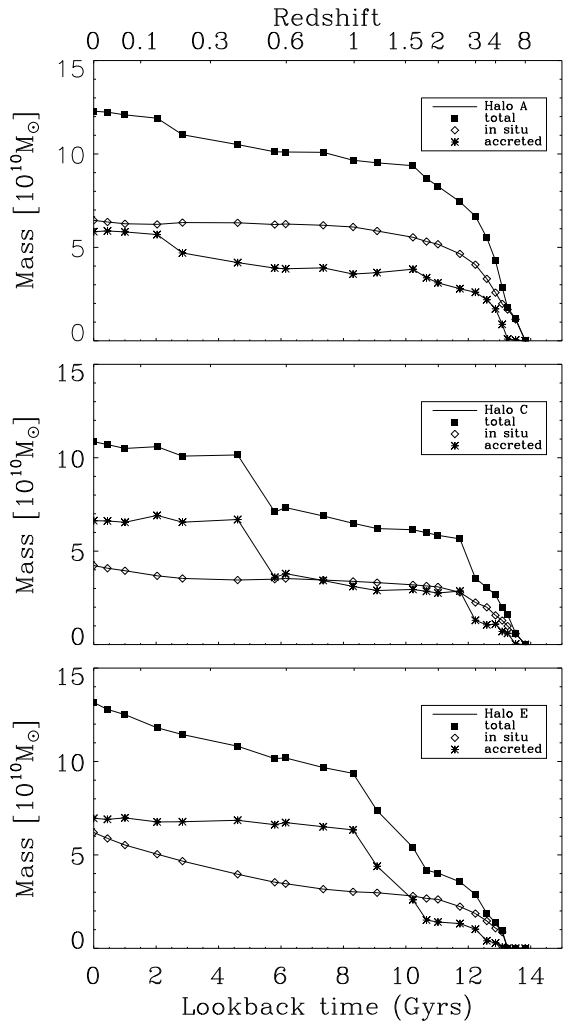}
  \caption{Accretion history for galaxies A,C and E of stars that have been 
added onto the central physical $30$ kpc (asterisk), stars that have formed in 
the galaxy in situ (open diamond) and the sum of the two (filled squares). 
\label{accrete_comb_col}}
\end{figure}

\section{Assembly histories for three galaxies} 
\label{ASSEMBLY}

The similarity in the star formation and assembly history  as well as in 
the present day properties of the $100^3$ and $200^3$ simulations for 
galaxy A encouraged us to believe that at a resolution  of $100^3$ 
particles (and the corresponding spatial and 
mass resolution) the simulations are numerically well enough resolved to run more
halos and find out whether the above results are typical. With current 
hardware and software it is difficult to run many halos at $200^3$ or 
even higher resolution.

The dark matter mass accretion history and the spin evolution 
of the three halos is shown in Fig. \ref{track_all}. We computed the mass 
within the virial radius defined by an enclosed mean overdensity 
of 200. By $z=1$ halos A and C have gained $\approx 40$\% of their final 
mass. Thereafter the halos continue to accrete dark matter and halo A has a
merger with a mass ratio of 6.5:1 at $z\approx0.6$, halo C has a 
major merger with a mass ratio of 3.5:1 at $z\approx0.8$. Halo E  
has already assembled more than 60\% of its mass by $z=1$ and undergoes an almost 
equal-mass merger at $z=1.6$. Thereafter the halos grow until $z=0$ without any
indication for a significant merger event. No halo experienced a major
merger within the last 4 Gyrs. 

The evolution of the halo spin parameter $\lambda^`$ defined as
\begin{eqnarray}
\lambda^{'} = \frac{J}{\sqrt{2} M_{\mathrm{vir}}v_{\mathrm{c}} r_{\mathrm{vir}}},
\end{eqnarray}
 
where $v_{\mathrm{c}}$ is the circular velocity at the virial 
radius $r{\mathrm{vir}}$ and $M_{\mathrm{vir}}$ is the virial mass 
\citep{2001ApJ...555..240B}, is shown in the bottom plot of Fig. 
\ref{track_all}. Halo E had gained significant spin during 
the $z=1.6$ major merger event which can clearly be identified by the sudden jump in 
$\lambda^`$. This merger induced increase of spin is in 
agreement with the findings of \citet{2002ApJ...581..799V} 
based on dark matter simulations. By $z=1$ the spin of halo E is larger 
than $\lambda^` =0.05$ and decreasing 
smoothly towards  $\lambda^` =0.03$ at $z=0$. In contrast, 
Halos A and C have low spin at $z=1$ and thereafter gain significant
spin during the merger events. The final spin values of all halos, 
$\lambda^`< 0.04$, is below the average predicted by dark matter simulations, 
however, their time evolution is consistent \citep{2002ApJ...581..799V} .  
 
In the top row of Fig. \ref{sfr_all} we show the accretion history of the baryonic 
matter subdivided into gas and stars within the virial radius of each halo. Halo
A has assembled most of its stellar mass by $z=1$ whereas halos C and E 
grow significantly at lower redshifts. By $z=0$ halo A has almost half of the 
baryons in gas with a ratio of stellar mass to gas mass of 
$M_{\mathrm{stars}}/M_{\mathrm{gas}} = 1.1$. Halo E, at a similar mass, has significantly 
more baryons in stars with $M_{\mathrm{stars}}/M_{\mathrm{gas}} = 1.6$. 
We derive the star formation rate (SFR) histories of the galaxies 
(bottom plot of Fig. \ref{sfr_all}) using the stellar ages at $z=0$ of all stars 
within $30$ kpc (again, we use this fiducial value as the stars in ellipticals 
typically can not be observed at larger radii). 
All galaxies start forming their stars in a burst at $z \approx 3-5$
with peak SFRs of $\approx  40-60 M_{\odot}$/yr and 
undergo an intense phase of merging (see the random walk of the halo spin parameters in 
Fig. \ref{track_all}) during which the SFRs decline exponentially towards $z=1$.
At $z=1$ the galaxies sustain star formation at rates of $3-7 M_{\odot}$/yr. 
For halos A and C the star formation rate drops to zero soon after their 
last merger event. About $2 Gyrs$ later small amounts of the previously heated gas 
is cooling at the center of these galaxies and the SFR is rising again to 
$\approx 2 M_{\odot}/$yr. Halo E, which did not undergo any significant 
merger after $z=1$ sustains an almost constant 
star formation rate until $z=0$.

\begin{figure}
  \centering
  \includegraphics[width=8cm]{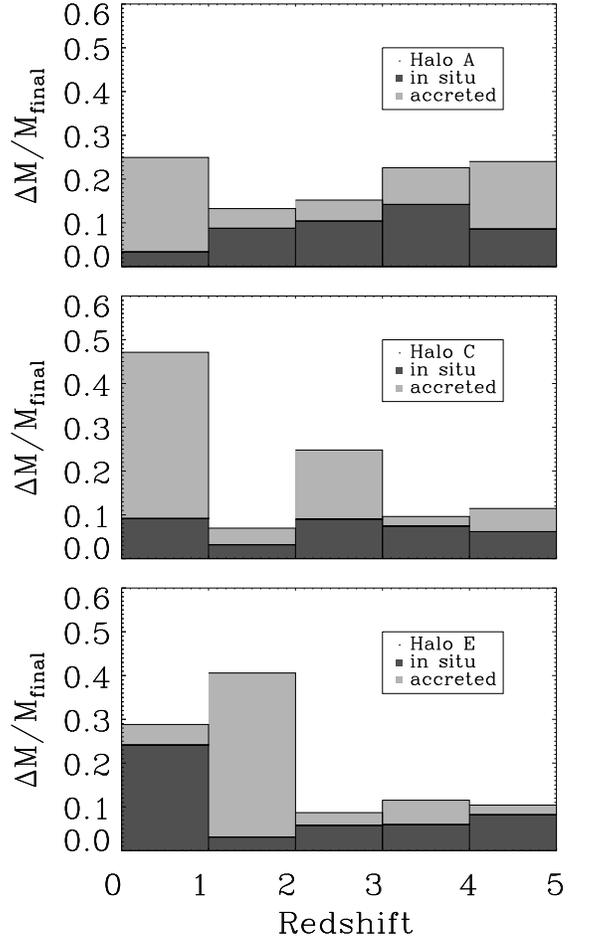}
  \caption{Accreted stellar mass and stars formed in situ as a function of 
redshift for the three galaxies. From $z=1\rightarrow 0$ galaxies A and C grow by 
gas poor mergers whereas galaxy E grows by in situ star formation. 
\label{Mass_frac_col_paper}}
\end{figure}

The above star formation histories were derived from the present day fossil record of
the stars in the galaxies. They indicate the time but not the location the stars were
born. The stellar mass of the galaxies is steadily increasing with 
time both by stars that were born from gas in situ in the galaxy as well as stars that 
have formed outside the galaxy and later on have been accreted. In
Fig. \ref{accrete_comb_col} we show the time evolution of stellar mass within a
fixed physical radius of 30kpc separated into stars that were born
within the galaxies and stars that have been accreted. By $z=1$ galaxy A
has already assembled 80\% of its final stellar mass when about 60\%
of the stars have formed inside the galaxy whereas 40\%
have been accreted by mergers. From $z=1$ to $z=0$ further mass assembly 
is dominated by accretion of stars which are accreted in 
small subunits. The most massive satellite is accreted at
$z\approx0.2$ and, after it has been stripped, has 10\% of the mass 
of the host galaxy. It had originally been accreted onto the halo in the 
merger event at $z=0.6$ that is described above. Galaxy C experiences a
major $\approx$ 3:1 merger at $z=0.6$. This merger can be identified as a gas
poor or dry merger as both progenitors were gas poor and the 
merger was not accompanied by a burst of star formation (see
Fig. \ref{sfr_all}). Galaxies A and C show on average an increase in 
mass by `dry' mergers of 24\% and only a few percent by in situ star formation 
from $z=1\rightarrow 0$. The situation is different for galaxy E which grows 
predominantly by in situ star formation from $z=1 \rightarrow 0$ (see Fig. 
\ref{Mass_frac_col_paper}).

\begin{figure}
  \centering
  \includegraphics[width=8cm]{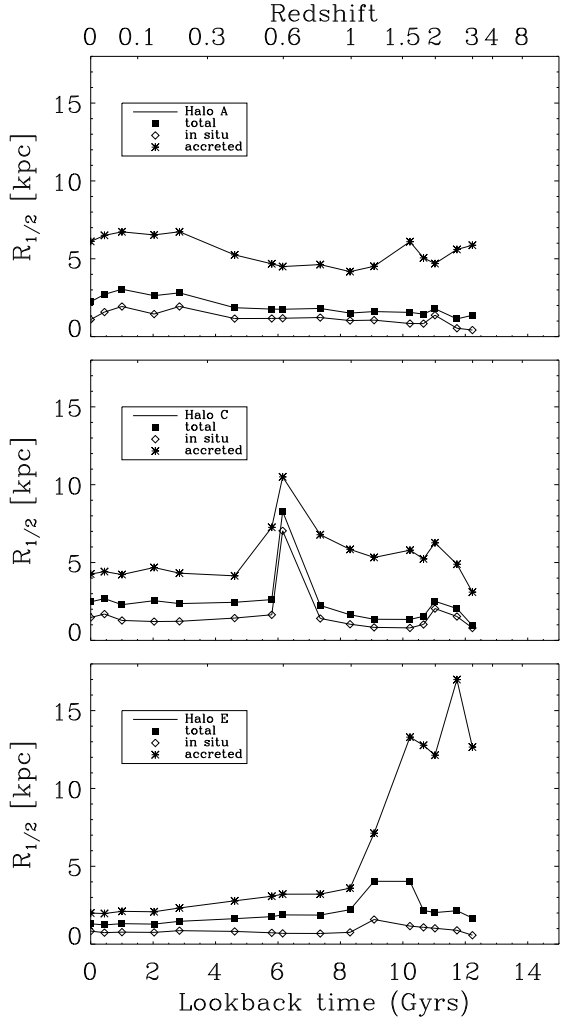}
  \caption{Time evolution of the projected spherical half-mass radius for galaxies A,C 
and E of all stars (filled squares) and the accreted (stars) 
and in situ stars (open diamonds) within a fixed radius of physical $30$ kpc. 
\label{accrete_half_mass_comb_col}}
\end{figure}

In Fig. \ref{accrete_half_mass_comb_col} we show the time evolution of the 
projected cylindrical half-mass radius (again within fixed physical 30kpc) for 
the galaxies as a whole and for the in situ and accreted population separately. 
The stars formed in situ have typical half mass radii of 1-2 kpc which changes 
only weakly with redshift. The accreted stars have significantly larger 
half-mass radii that lead to an increase of the total half-mass radius for galaxies 
A, C and E as the accreted mass is increasing towards the present time. The situation 
is different for galaxy E: the in situ half mass radius is constant as well, however, the 
mass in in situ stars is increasing. As a result the half mass radius of the galaxy 
is decreasing similar to adiabatic contraction \citep{1986ApJ...301...27B,2002ApJ...571L..89J}.

\begin{table*}
\caption{Galaxy properties for $r < r_{\mathbf{vir}}$ }             
\label{all_vir}      
\centering          
\begin{tabular}{c| c c c c c c c c |c c c c c c}   
\hline\hline       
Galaxy & $M_{\mathbf{vir}}$\tablenotemark{(a)} & $M_{\mathbf{stars}}$ &$M_{\mathbf{gas}}$ & $M_{\mathbf{dark}}$ &  $r_{\mathbf{vir}}$ \tablenotemark{(b)}&  $v_{\mathbf{max}}$ \tablenotemark{(c)}& $<\Lambda^`>$ \tablenotemark{(d)}& $f_{hot}$ \tablenotemark{(e)}  &  $m_{\mathbf{stars}}$ \tablenotemark{(f)} &$m_{\mathbf{dark}}$& $\epsilon_{\mathbf{stars}}$ \tablenotemark{(g)}& $\epsilon_{\mathbf{dark}}$  \\ 
\hline                    
   A & 229 &  23.8 & 20.9 & 184 & 426 & 270 & 0.022 $\pm$ 0.012 & 0.99 & 10.5 & 82.5 & 0.25 & 0.5\\  
   C & 158 &  18.0 & 13.2 & 127 & 377 & 257 & 0.034 $\pm$ 0.012 & 0.99 & 8.5  & 69   & 0.25 & 0.5 \\
   E & 171 &  21.3 & 13.1 & 137 & 387 & 307 & 0.039 $\pm$ 0.010 & 0.98 & 8.5  & 69   & 0.25 & 0.5 \\
\hline                  
\end{tabular}
\tablecomments{(a) Total masses $M$ in$10^{10}M_{\odot}$; (b) Virial radius in kpc; 
(c) Maximum circular velocity in km/s; (d) Time averaged halo spin parameter;  
(e) Fraction of hot gas ($T > 10^5$ K); 
(f) Particle masses $m$ in $10^{5}M_{\odot}$; (g) Gravitational softening lengths in kpc}
\end{table*}

\begin{table*}
\caption{Galaxy properties for $r <$ 30kpc}             
\label{all_30}      
\centering          
\begin{tabular}{c c c c c c c c c c c c c c c c}     
\hline\hline       
Galaxy & $M_{stars}$ \tablenotemark{(a)}& $M_{gas}$ & $f_{hot}$ \tablenotemark{(b)}& $r_{1/2}$\tablenotemark{(c)}& $<$age$>_*$ \tablenotemark{(d)}&$M_{\mathbf{B}}$\tablenotemark{(e)} &  $SB_{\mathbf{eff}}$ \tablenotemark{(f)}& $r_{\mathrm{eff}}$\tablenotemark{(g)}&  $f_{\mathrm{DM}}$\tablenotemark{(h)} & $\epsilon$\tablenotemark{(i)} &  $a_4$ \tablenotemark{(j)}& $\sigma_{0}$ \tablenotemark{(k)}& $v_{\mathbf{maj}}$\tablenotemark{(l)}&  $<H_3>$ \tablenotemark{(m)} & T\tablenotemark{(n)} \\ 
\hline                    
   A & 12.3 & 0.64 & 0.95& 2.2  & 11.5 & -20.18 & 20.5 & 2.7 & 0.38 &0.18 & 0.6 & 153 & 43 & -0.05 & 0.08\\  
   C & 10.8 & 0.82 & 0.90 &2.5 & 11.2 & -20.45 & 20.3 & 2.8 & 0.35 &0.21 & 1.1 & 145 & 84 & -0.09 & 0.77\\
   E & 13.2 & 0.84 & 0.88 & 1.4 &  9.8 & -21.06 & 19.7 & 1.9 & 0.18 & 0.5 & 2.5 & 180 & 225& -0.05 & 0.13\\
\hline                  
\end{tabular}
\tablecomments{(a) Total masses $M$ in$10^{10}M_{\odot}$; (b) Fraction of hot gas ($T > 10^5$ K); 
(c) Projected stellar half-mass radius; (d) Average stellar age at z=0;  
(e) Absolute magnitude in B-band; 
(f) Effective surface brightness in B-band in mag arcsec$^{-2}$; (g) Fitted effective radius; 
(h) fraction of dark matter within $r_{\mathrm{eff}}$; 
(i) Ellipticity at $r_{\mathrm{eff}}$; (j) Isophotal shape parameter; 
(k) Central stellar velocity dispersion in km/s; 
(l) Major axis velocity in km/s; (m) LOSVD assymetry parameter; (n) Triaxiality parameter}

\end{table*}

\section{Present day properties}

\begin{figure}
  \centering
  \includegraphics[width=8cm]{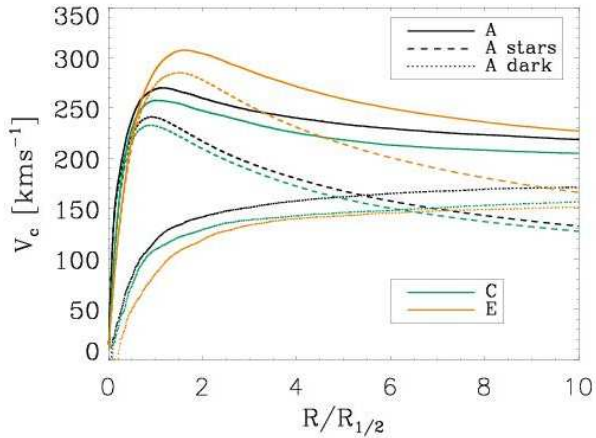}
  \caption{Circular velocity curves for all galaxies versus radius normalized to their 
projected stellar half-mass radius. All galaxies are dominated by luminous matter inside 
two half-mass radii. \label{rotcurve}}
\end{figure}

\begin{figure}
\centering 
\includegraphics[width=8cm]{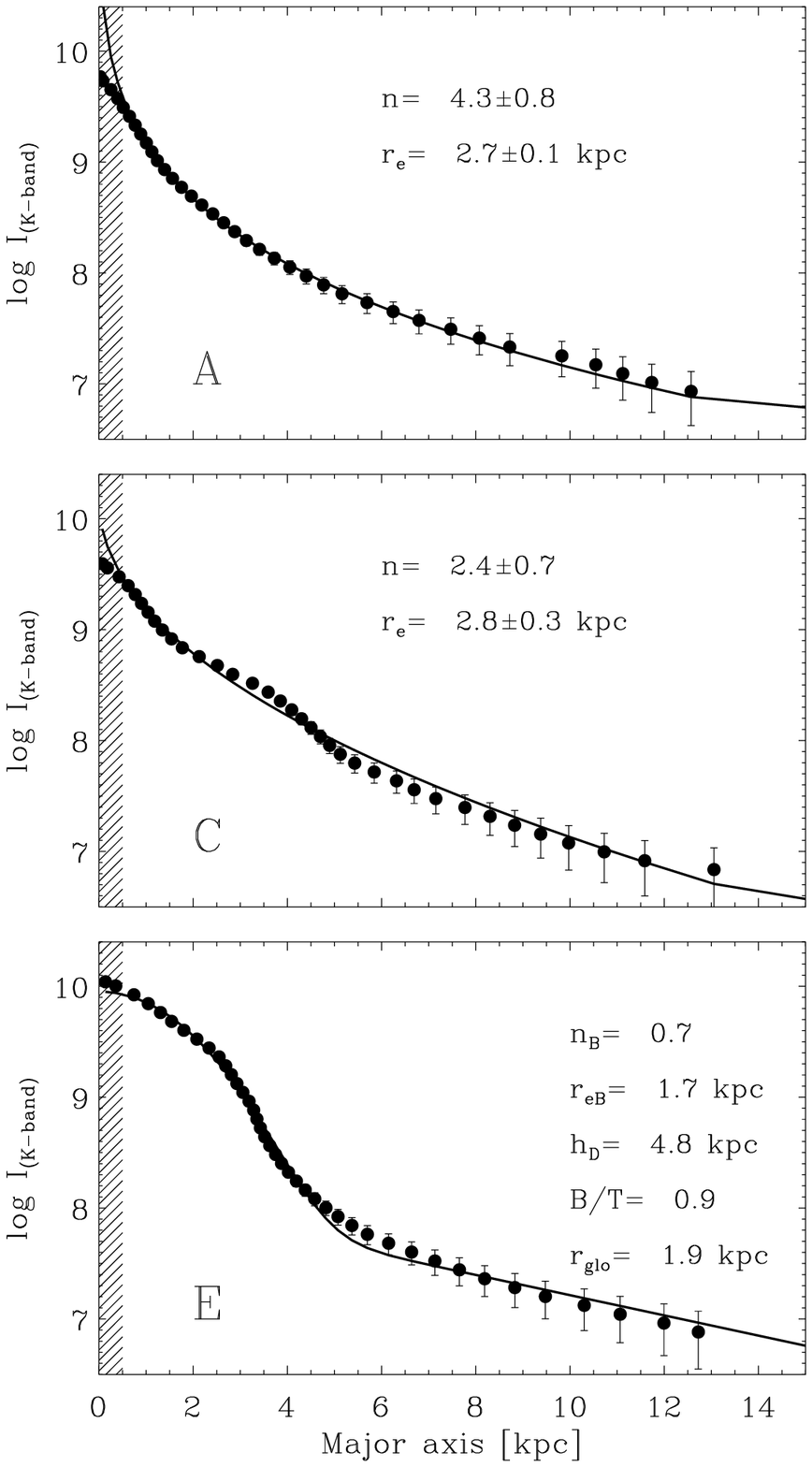}
\caption{K-band surface brightness profiles for galaxies A,C, and E. The profiles inside 
$\approx 2$ softening lengths (shaded area) were excluded from the fits as they are 
likely to be influenced by force softening. The error bars
are given by the Poissonian error of the particles in the fitted regime. The errors 
for the S\'ersic index $n$ and the effective radius $r_{\mathrm{e}}$ have been determined 
by the bootstrapping method. Galaxies A and C can be fitted well by a single 
S\'ersic function whereas galaxy E can only be fitted by a two component fit (see text). 
\label{kmag_all}}
\end{figure}

We have summarized the properties inside the virial radius of the galaxies in Tab. 
\ref{all_vir}. In Fig. \ref{rotcurve} we show the circular velocity curves for all 
galaxies normalized to their projected stellar half mass radii. These seem 
very reasonable, in marked contrast to the results of e.g. \citet{2003ApJ...590..619M}.  
The galaxies are dominated by luminous matter inside $10$ kpc and the fraction of 
dark matter within the half-mass  radii is $0.18 < f_{\mathrm{DM}} \approx 0.38$. 
This is consistent with most findings based on observations and dynamical 
modeling that early-type galaxies are dominated by luminous matter in their inner parts 
\citep{1982MNRAS.201..975E,1997ApJ...488..702R,2003Sci...301.1696R,2004ApJ...611..739T,
2005MNRAS.360.1355T}. However, dynamical modeling of observed ellipticals 
also indicates that dark matter becomes dominant beyond 2-3 half-mass 
radii whereas for the simulated galaxies A,C, and E dark matter becomes 
dominant at a factor of two larger radii.   

The properties of the central stellar component of the galaxies 
are summarized in Tab. \ref{all_30}.
At present galaxies A and C have very old stellar population with mean ages
of about $11\;\rm{Gyrs}$, galaxy E which had significant recent star formation 
is more than one Gyr younger. The final stellar masses within 30kpc are in the range of 
$10.8 \times 10^{10} M_{\odot} < M_{*} < 13.2 \times 10^{10} M_{\odot}$. 
We have computed the luminosities of the galaxies using the \citet{2003MNRAS.344.1000B} 
spectro-photometric models assuming a Salpeter IMF and solar metalicity.  
As all galaxies had recent star formation on various levels (see Fig. \ref{sfr_all}), 
they are shifted blue-ward from the color magnitude relation of cluster 
ellipticals (see e.g. \citealp{2005ApJ...619..193M}). However, for galaxies A and C the 
effect is not very strong and any process that would  suppress central star 
formation within the last $10^8$ yrs would shift the galaxies 
(with the exception of galaxy E) to the observed relation. 
Higher numerical resolution might result in a similar effect, as galaxy A at $200^3$ 
resolution falls right on the color-magnitude relation.  However, it is not unusual 
for field ellipticals (E+A phenomenon) to actually show residual star formation  
\citep{2004MNRAS.355..713B,2006astro.ph..1036S}. 

The simulations had sufficient resolution 
(a few $10^5$ particles within 30$\rm kpc$) to analyze the kinematic and 
photometric properties in more detail. In Fig. \ref{kmag_all} we 
show the edge-on K-band surface brightness profiles of the galaxies and the 
best fitting S\'ersic function or the best two component S\'ersic+exponential fit
\citep{2006MNRAS.369..625N}. With S\'ersic indices of $n = 4.3$ and $n = 2$
and effective radii of $2.7 \;\rm kpc$ and  $2.8 \;\rm kpc$ galaxies 
A and C can best be fitted by a single S\'ersic function and have a 
very reasonable concentration in good agreement with observed intermediate 
mass ellipticals \citep{2004ApJ...600L..39T}. 
Galaxy E can not be fitted with a single component. It is dominated by a central 
'bulge' component with a S\'ersic index of $n_B=0.5$ and a size of $r_{eB}= 1.7$kpc  
and an outer exponential with a scale length of $h_d=4.8kpc$. The bulge-to-total ratio 
is $B/T = 0.9$. In fact the central component is a compact exponential disk seen 
edge-on. 

\begin{figure}
\centering 
\includegraphics[width=8cm]{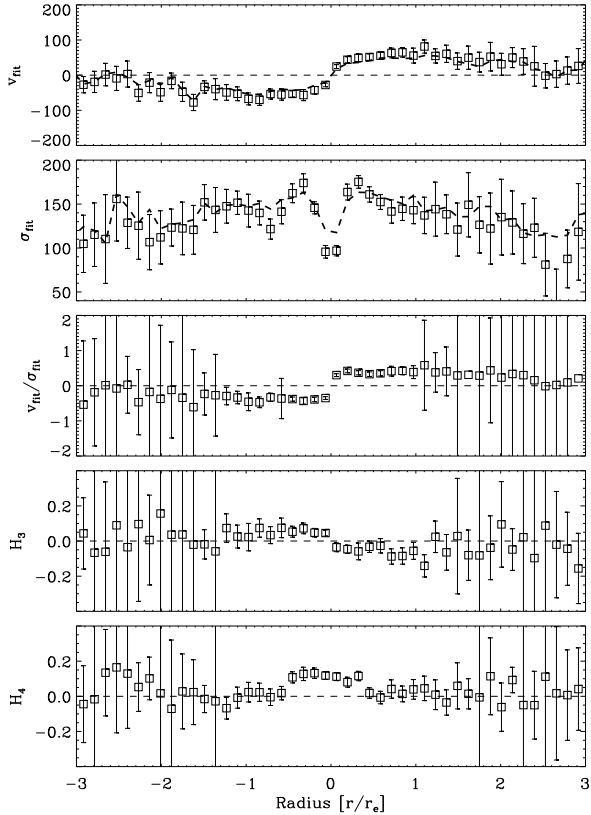}
\caption{Analysis of the LOSVD of galaxy A as measured along
   a slit aligned with the major axis of the moment of inertia tensor
   of the main stellar body (open squares). The fitted local velocity
   $v_{\mathrm{fit}}$, local velocity dispersion $\sigma_{\mathrm{fit}}$, 
    $v_{\mathrm{rot}}/\sigma_{\mathrm{fit}}$, $H_3$, and $H_4$ are
   plotted versus radius. The dashed lines show the
   true line-of-sight velocity and dispersion, respectively. The
   individual error-bars were derived by
   bootstrapping.
\label{losvd_cosmo_boot_A}}
\end{figure}

\begin{figure}
\centering 
\includegraphics[width=8cm]{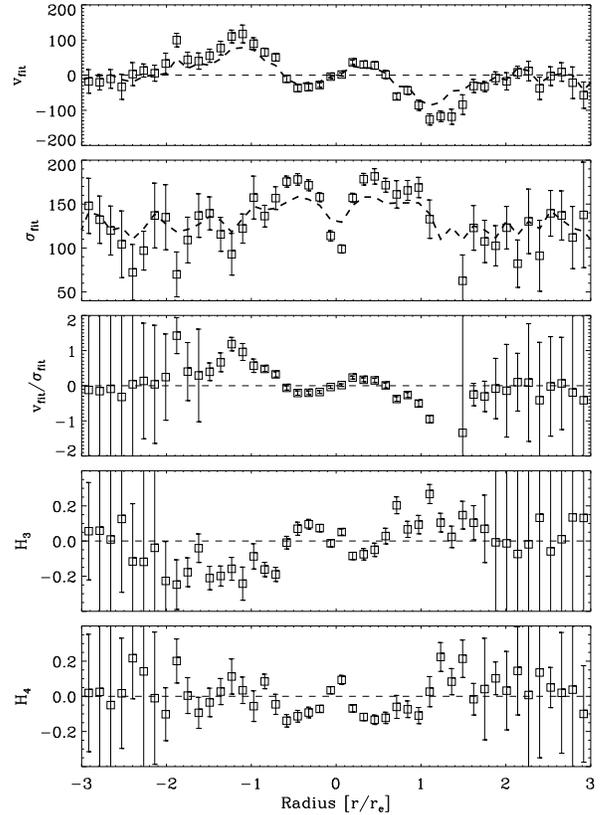}
\caption{Sam as Fig. \ref{losvd_cosmo_boot_A} for galaxy C. Note that this galaxy has 
a counter-rotating core with a radius of $\approx 0.5 r_{\mathbf{e}}$.
\label{losvd_cosmo_boot_C}}
\end{figure}

We have investigated the LOSVDs of the galaxies by binning the velocities 
on a slit along the apparent long axis of each projected galaxy. 
Subsequently we parameterized deviations from the Gaussian shape using 
Gauss-Hermite basis functions \citep{1993MNRAS.265..213G,1993ApJ...407..525V}.  
The kinematic parameters of each profile ($v_{\mathrm{fit}}$, 
$\sigma_{\mathrm{fit}}$, $h_3$, $h_4$) were then determined by least squares fitting
as in \citet{2006MNRAS.372..839N}. As an example we show the fitted kinematical profiles 
for the edge on projection  of galaxies A and C in Figs. \ref{losvd_cosmo_boot_A} and 
\ref{losvd_cosmo_boot_C}. As they 
are affected by higher order moments we show the real line-of-sight velocity and 
dispersion in addition (dashed lines).  
Both galaxies show significant rotation. Galaxy C has a counter-rotating core that has 
formed during the major merger event. The velocity 
dispersion profiles are peaked towards the central region and drop in the outer parts. 
The drop at the very center is caused by stars that have formed late from the cooled halo 
gas. A similar behavior is seen in some field ellipticals like, 
e.g. NGC2768 \citep{2004MNRAS.352..721E} and could have the same origin. 
The asymmetry of the LOSVD measured by $h_3$ is always anti-correlated to the 
line-of-sight velocity, whereas the LOSVD is more peaked than a Gaussian at the center, 
indicated by a positive value of $h_4$. The errors have been computed applying the 
bootstrap algorithm. Even for the higher order moments they are reasonably small within 
the effective radius. 

\begin{figure}
\centering 
\includegraphics[width=8cm]{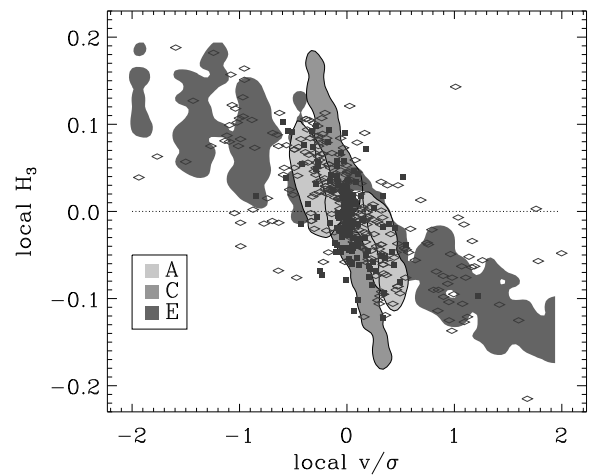}
\caption{Local correlation between $h_3$ and
$v_{\mathrm{fit}}/\sigma_0$ for galaxies A,C, and E for 500 radom projections. 
90\% of the data points cover the regions indicated by the shaded areas. The observed values
for disky (open diamonds) and boxy ellipticals (filled boxes) are overplotted 
\citep{1994MNRAS.269..785B}. 
\label{overlay_h3_cosmo_pub}}
\end{figure}

In Fig. \ref{overlay_h3_cosmo_pub} we show the location of 500 random projections of 
galaxies A,C, and E in the $v_{\mathrm{fit}}/\sigma_{\mathrm{fit}}$-$h_3$ plane. The shaded 
area indicate the 90\% probability to find projected galaxy with the given properties. 
All projected galaxies show steep leading wings in their LOSVD 
(anti-correlated $h_3$ and $v_{\mathrm{fit}}$). 
The location of the galaxies is in good agreement with observations of elliptical galaxies 
indicated by the symbols in Fig. \ref{overlay_h3_cosmo_pub} 
\citep{1994MNRAS.269..785B,2006MNRAS.372..839N}. Galaxy E is the fastest 
rotator with $v_{\mathrm{fit}}/\sigma_{\mathrm{fit}} \approx 2$.  
 
\begin{figure*}
\centering 
\includegraphics[width=16cm]{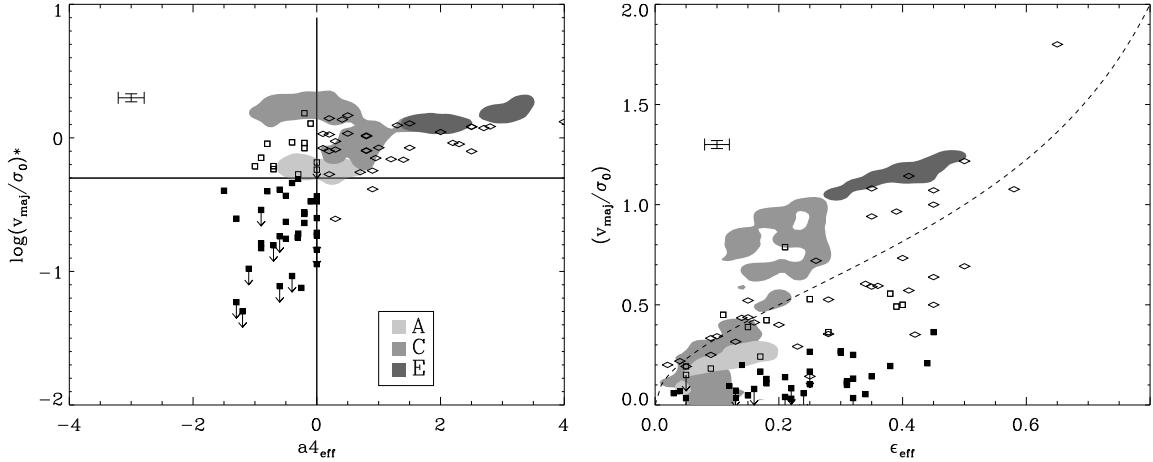}
\caption{{\it Left panel}: Isophotal shape $a4_{\mathrm{eff}}$ versus anisotropy 
parameter $(v_{\mathrm{maj}}/\sigma_0)^*$. The contours indicate the 90\% 
probability to find a simulated galaxy (seen from 500 random viewing angles) 
in the enclosed area. The horizontal line at 
$(v_{\mathrm{maj}}/\sigma_0)^* =0.5$ separates the anisotropic boxy 
from the more rotationally flattened disky and boxy ellipticals. 
{\it Right panel}: Ellipticity $\epsilon_{\mathrm{eff}}$ versus the ratio of 
major axis rotation velocity and central velocity dispersion $(v_{maj}/\sigma_0)$.   
The data for observed disky (diamonds), very anisotropic 
boxy (filled boxes) and less anisotropic boxy (open boxes) ellipticals have 
been kindly provided by Ralf Bender. The properties of the galaxies are largely 
consistent with rotationally flattened disky and boxy ellipticals. 
\label{overlay_a4vsig_vsell_cosmo_pub}}
\end{figure*}

We determined the characteristic ellipticity $\epsilon_{\mathrm{eff}}$, 
the isophotal shape $a4_{\mathrm{eff}}$ ($a4_{\mathrm{eff}} <0$: 
boxy isophotes; $a4_{\mathrm{eff}}>0$: disky isophotes) as well as the central 
velocity dispersion $\sigma_0$, the maximum rotation 
velocity $v_{\mathrm{maj}}$, and the triaxiallity 
parameter $T$ as in \citet{2003ApJ...597..893N}. 
In Fig. \ref{overlay_a4vsig_vsell_cosmo_pub} we indicate the location of the 
simulated galaxies A,C, and E in the plane defined by the anisotropy parameter 
$v_{\mathrm{maj}/\sigma_0}^*$ and the isophotal shape $a4_{\mathrm{eff}}$. 
We have computed the parameters for 500 random projections of each galaxy and show 
the area enclosing the 90\% most likely observed data points. For comparison we show the 
observations for disky (diamonds) and boxy (squares) ellipticals galaxies. All galaxies
are similar to isotropic rotators (defined here as $\log(v_{\mathrm{maj}/\sigma_0}^*) 
> 0.7$ ). Galaxy A is only marginally disky, Galaxy C can be very disky but also 
has some boxy projections, both are in good agreement with observations of disky and boxy 
isotropic ellipticals.  Galaxy E has the largest values for $v_{\mathrm{maj}/\sigma_0}^*$ 
and is extremely disky. It also show the largest values for rotational support 
$v_{\mathrm{maj}/\sigma_0}>1$ (Fig. \ref{overlay_a4vsig_vsell_cosmo_pub}, right panel) 
whereas galaxy C shows smaller values $v_{\mathrm{maj}/\sigma_0} \approx 0.7$ and 
ellipticites as large as $\epsilon = 0.5$. Galaxy a has the lowest values of 
$v_{\mathrm{maj}/\sigma_0} \approx 0.2$ and is nearly round at$\epsilon = 0.2$.  

In summary the high star-formation rate, the exponential surface brightness distribution, 
the kinematics and the isophotal shape of galaxy E makes it more similar to an early-type 
spiral galaxies. Galaxies A and C do resemble $M_*$ elliptical galaxies. 
They follow the Fundamental Plane ($[\sigma_0,SB_e,r_e]=[153 \rm\,
{km/s}, 20.5 \rm mag\,arcsec^{-2}, 2.7 \rm\,kpc]$; $[\sigma_0,SB_e,r_e]=[145 \rm\,
{km/s}, 20.3 \rm mag\,arcsec^{-2}, 2.8 \rm\,kpc]$) \citep{1992ApJ...399..462B}
and are consistent with the observed $n-M_B$ and the $n-\sigma$ 
correlation \citep{2004ApJ...600L..39T}.


\section{Discussion}
We have followed the formation and evolution of three $\approx M_*$ field 
galaxies using numerical simulations 'ab initio' from cosmological initial conditions at 
high redshift. All galaxies start to form when cold gas that has collapsed 
in subunits is effectively consumed by star formation. At early times
the remaining diffuse gas is predominantly heated by shocks. In
general the stars galaxies assemble both by in situ star formation as
well as major/minor mergers and accretion. During the early formation
phase at $2<z<8$ the assembly of the galaxies is dominated by mergers
of gas rich subcomponents and  in situ star formation. Although some
fraction of the stars is accreted this phase has  the characteristics
of a dissipative collapse. Thereafter stellar accretion or minor/major
mergers become more important and , for early-type galaxies, tends to
dominate for $z < 1$.  

After z=1 the further assembly of the galaxies varies from halo to halo. 
The evolution of one galaxy is dominated by gas infall and in situ star formation at 
rates of $5-8 M_{\odot}$/yr. It develops a massive, however too
compact, stellar disk and has  properties more similar to early-type
disk galaxies. 

Two other galaxies have present 
day properties more similar to elliptical galaxies and their assembly is dominated by minor 
and major predominantly stellar mergers. One of the galaxies grows by minor mergers 
and does not experience  a major merger after its formation phase. At redshift $z \approx 1$ 
it has formed a massive stellar spheroid (see e.g. \citealp{2005ApJ...618...23N}) 
which resembles observed massive evolved galaxies and their inferred formation 
histories \citep{2000A&A...362L..45D}. The system evolves into an old
stellar system surrounded by a halo of hot gas at the present day. The
second early-type galaxy had a major merger at z=0.6 that can be
classified as a gas poor or 'dry' merger as it was not accompanied by
a strong  burst of star-formation. From those two examples we find
that $\approx$ 15\% - 35\% of the final  
stellar mass was assembled by accretion of stars that have formed
outside the galaxy (see \citealp{2006MNRAS.370..902K}). 
This increase will contribute to the evolution of the luminosity function of 
elliptical galaxies from $z=1\rightarrow 0$ in addition to a fading population of 
blue galaxies and mergers \citep{2004ApJ...608..752B,2004ApJ...608..742D,2005ApJ...620..564C,
2005astro.ph..6044F}. The size evolution 
of the simulated early-type galaxies is determined by the accreted stars. While 
the half-mass radius of the stars formed in-situ remains almost constant independent 
of redshift at $r_{1/2}= 1-2$ kpc the accreted stars create an envelope whose 
half-mass radius is increasing with decreasing redshift (see
\citealp{2006ApJ...648L..21K}, for a semi-analytical approach). Interestingly, a significant 
fraction of the stars at the final half-mass radius of the galaxies were not born there 
but accreted. This process might provide a good 
explanation for a possible size evolution of early-type galaxies 
\citep{2006ApJ...650...18T,2006astro.ph.10241L} without changing the
fundamental parameter relations which are determined by the central
properties of the galaxies and should have been established during the
early formation phase.

At the present day most investigated kinematical and photometric properties 
(see also \citealp{2003ApJ...596...47S}) are consistent with observed rotating 
disky and boxy early-type galaxies at similar luminosities. In particular, the 
concentration is in good agreement with observations. We believe that the discrepancy with 
\citet{2003ApJ...590..619M} is due to the different star 
formation algorithm which in their case allows predominantly gaseous subunits 
to merge and develop a high concentration before forming stars. In our simulations 
gas is converted into stars efficiently in subunits which merge at early times, 
prior to z=2, and so the morphology of the systems is elliptical if its late accretion is 
dominated by stars. Interestingly, the early-type galaxy with the late major merger develops a 
counter rotating core which previously only have been successfully simulated in isolated 
equal-mass disk-disk mergers \citep{1991Natur.354..210H,2006astro.ph..6144J} and are 
supposed to be signposts of merger events.

Feedback from supernovae and/or AGN, which was not included in this simulation, 
would provide additional heat sources but appear neither vital to stop the star formation  
and produce a massive red galaxy at higher redshifts nor to produce a present 
day stellar system with consistent kinematical properties. Theoretical
arguments (see {\it e.g.} \citealp{1986ApJ...303...39D};  
\citealp{2000MNRAS.317..697E}) suggest that stellar feedback rapidly
becomes ineffective in systems with velocity dispersions greater than
100 km/s. However, AGN feedback may be important in galaxies with high
velocity dispersions \citep{1998A&A...331L...1S,1999MNRAS.308L..39F,2001ApJ...551..131C}. 
In our case AGN feedback might have been able to prevent further star
formation at low redshift by heating the cold gas at the center of the
galaxy (see e.g. \citealp{2005ApJ...620L..79S,2005MNRAS.358..168S}).  
On the other hand many field ellipticals show the related E+A phenomenon which seems 
to be suppressed in the centers of clusters as any residual cold gas 
will be blown away \citep{1993AJ....106..473C}. For more massive
systems than the ones studied here it has been argued that feedback from AGN might 
have a bigger impact and be responsible for the sharp cutoff at the bright end of the 
luminosity function of ellipticals \citep{2005MNRAS.356.1155C}. However, only full 
cosmological simulations at high resolution of the formation of more massive systems will 
help to fully understand the importance of stellar and AGN feedback on the formation and 
evolution of elliptical galaxies.   

In a hierarchical universe most galaxies are expected 
to experience mergers but it has been argued for a long time that dissipation 
is essential to explain properties of elliptical galaxies and in particular 
intermediate mass ellipticals (\citealp{1989ApJ...342L..63K} and references therein; 
\citealp{1996ApJ...464L.119K,2006MNRAS.372..839N}). The cosmological formation 
processes presented here combines characteristics of both stellar mergers and 
dissipational collapse/mergers whereas the latter dominates the early formation phase.
It naturally meets many observational constraints on the evolution 
of massive spheroidal galaxies (e.g. \citealp{2005ApJ...633..174T,2005ApJ...621..673T,
2005ApJ...631..145V}). It might be equally or even more important than 
major mergers of disk galaxies \citep{2006ApJ...641...21R,2006ApJS..163....1H} 
which can not reproduce the properties the most massive ellipticals \citep{2003ApJ...597..893N,2006astro.ph..7446C} 
whose evolution could be driven by mergers of evolved ellipticals \citep{1977ApJ...217L.125O,2003ApJ...597L.117K,2006ApJ...640..241B,2006ApJ...636L..81N,2006MNRAS.369.1081B}.

\begin{acknowledgements}
We thank Volker Springel for making GADGET2 available prior to
publication and Ralf Bender and Andy Fabian for discussions 
about the manuscript.We also thank the anonymous referee for valuable
comments on the manuscript.  

\end{acknowledgements}

\end{document}